\documentstyle[12pt,epsf]{article}

\newcommand{\la}[1]{\label{#1}}
\newlength{\numlen}

\newlength{\indexlength}

\newcommand{\be}{\begin{equation}}
\newcommand{\ee}{\end{equation}}
\newcommand{\ba}{\begin{eqnarray}}
\newcommand{\ea}{\end{eqnarray}}

\newcommand{\eq}{eq.~}

\newcommand{\fig}{fig.~}

\newcommand{\nr}[1]{(\ref{#1})}

\newcommand{\h}{{\hspace{0.5 cm}}}

\begin{document}
% Title

\begin{titlepage}
\mbox{}\hfill FSU-SCRI-94-123\\
\mbox{}\hfill AZPH-TH/94-26\\
\mbox{}\hfill IUHET-288\\
\mbox{}\hfill UUHEP-94/5\\
\begin{centering}
\vfill

{\bf THERMAL PHASE TRANSITION IN MIXED ACTION SU(3) LATTICE GAUGE THEORY
AND WILSON FERMION THERMODYNAMICS}

\vspace{1cm}
T.~Blum$^a$, C.~DeTar$^b$, Urs M.~Heller$^c$,
Leo~K\"arkk\"ainen$^a$, K.~Rummukainen$^d$, and D.~Toussaint$^a$

\vspace{1cm}
{\em $^a$Department of Physics, University of Arizona,\\
Tucson, AZ 85721, USA\\}

\vspace{0.3cm}
{\em $^b$Department of Physics, University of Utah,\\
Salt Lake City, UT 84112, USA\\}

\vspace{0.3cm}
{\em $^c$SCRI, The Florida State University,\\
Tallahassee, FL 32306-4052, USA\\}

\vspace{0.3cm}
{\em $^d$Department of Physics, Indiana University,\\
Bloomington, IN 47405, USA\\}

\vspace{1.5cm}
{\bf Abstract}

\end{centering}

\vspace{0.3cm}\noindent
We study the thermal phase diagram of pure SU(3) lattice gauge theory with
fundamental and adjoint couplings. We improve previous estimates of the
position of the bulk transition line and determine the thermal
deconfinement transition lines for $N_t=2,4,6,$ and 8. The endpoint of the
bulk transition line $(\beta_f, \beta_a)=(4.00(7), 2.06(8))$ improves upon
earlier estimates obtained using smaller lattice sizes. For $N_t > 4$ the
deconfinement transition line splits cleanly away from the bulk transition
line. With increasing $N_t$ the thermal deconfinement transition lines
shift to increasingly weaker coupling, joining onto the bulk transition
line at increasingly larger $\beta_a$ in a pattern consistent with the
usual universality picture of lattice gauge theories. We also discuss the
possible consequences of an induced adjoint term from the fermionic
determinant and determine this induced term numerically with a
microcanonical demon method for two flavors of dynamical Wilson fermions.

\vfill
\end{titlepage}

\section{Introduction}

\begin{figure}[htb]
\begin{center}
\leavevmode
\epsfysize=360pt
%\epsfbox{phas_diag_T0.eps}
\epsfbox[90 40 580 490]{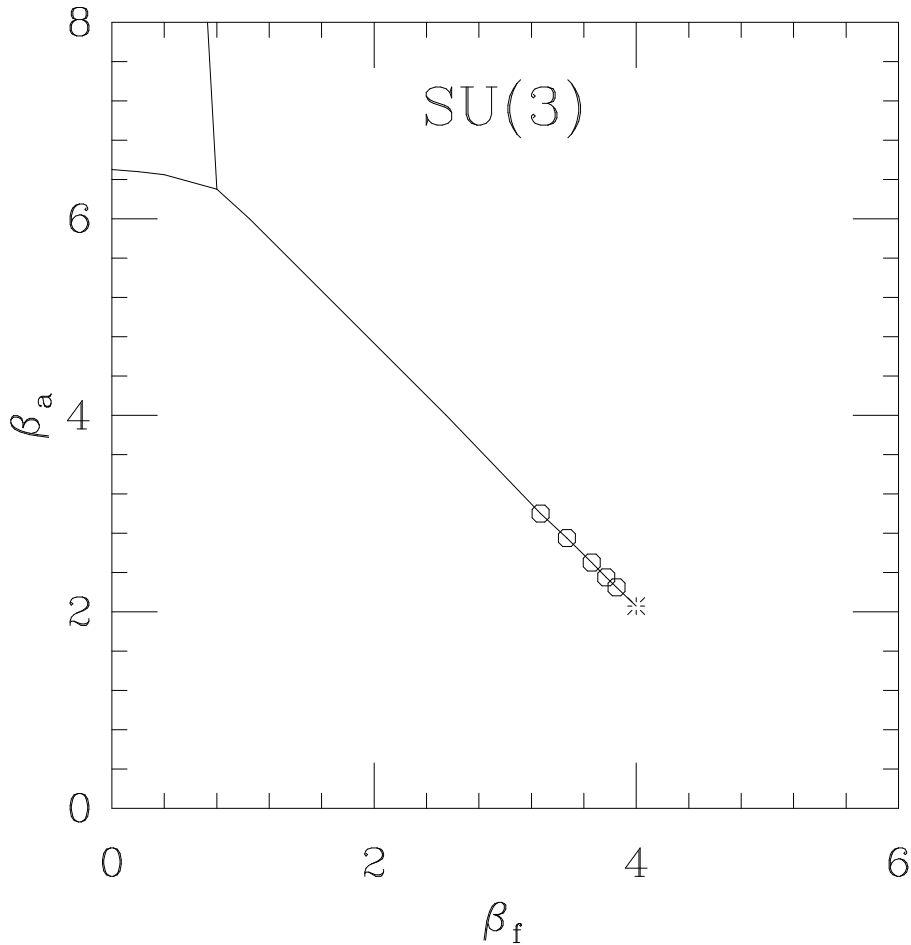}
%\epsffile{phas_diag_T0.eps}
\end{center}
\caption{The phase diagram for the fundamental--adjoint system.}
\label{phas_diag_T0}
\end{figure}

The phase diagram of fundamental--adjoint pure gauge systems, shown in
Fig.~\ref{phas_diag_T0}, is considerably more complicated than the one with
only a fundamental coupling. Early studies found that it has first order
(bulk) transitions in the region of small $\beta_f$
\cite{Greensite81,Bhanot82}. For SU(3) the purely adjoint system has a
transition at $\beta_a = 6.5(3)$. For $\beta_a \to \infty$ the system
reduces to a $Z_3$ gauge theory with a transition at $\beta_f = 0.67$. In
the $\beta_a$,$\beta_f$ plane these transition points are extended to lines
that merge into a single terminating line segment as shown. As discussed in
Sec.~\ref{phase_diag}, we place the endpoint at $(\beta_f,
\beta_a)=(4.00(7), 2.06(8))$, considerably different from old results
obtained on small lattices \cite{Bhanot82}. We also find a roughly straight
line of bulk crossovers extending beyond the endpoint.

The non-trivial phase structure in the fundamental--adjoint plane, and in
particular the critical endpoint of the transition line pointing toward the
fundamental axis, has been shown to be associated with, or even
responsible, for the dip in the discrete $\beta$-function of the theory
with standard Wilson (fundamental) action, which occurs in the region where
the bulk crossover line crosses the fundamental axis. The bulk transition
might also mask the thermal deconfinement transition, at least for lattices
with small temporal extent $N_t$. Indeed, from early simulations by
Batrouni and Svetitsky \cite{Batrouni84} and by Gocksch and Okawa
\cite{Gocksch84} we know that for pure gauge SU(4) theory, for which the
bulk transition crosses the fundamental axis, the $N_t = 2$ temperature
transition joins the bulk line, but that for $N_t=4$ the bulk line and the
thermal line exist separately.

In a recent paper Gavai, Grady and Mathur showed that at nonzero
temperature the deconfinement transition in pure gauge SU(2) survives for
positive adjoint couplings and is connected to the end point of the bulk
transition line \cite{Gavai94}. They were not able to distinguish the first
order bulk line from the $N_t = 4$ and 6 thermal transition lines in their
simulations. Universality in the continuum limit requires that as $N_t$ is
increased, the thermal phase transition line shifts to weaker coupling, so
that approaching the zero-temperature weak-coupling limit along any line in
the fundamental--adjoint coupling plane leads to the same low temperature,
confined theory. Thus the thermal phase transition lines could not remain
anchored to a bulk transition line for higher $N_t$. More recently Mathur
and Gavai~\cite{Gavai94b}  raised doubts about the very existence of a bulk
phase transition in SU(2), characterizing even the first-order portion of
the line as a thermal deconfinement transition that is displaced toward
weak coupling with increasing $N_t$. Clarification is obviously needed.

Puzzled by the finding of Ref.~\cite{Gavai94}, but also motivated by some
unexpected results in the study of the high temperature behavior of full
QCD with Wilson fermions~\cite{wilson_woes} that might be explained by the
fact that the fermion determinant induces, among others, an effective
adjoint coupling, we decided to study pure SU(3) gauge theory with a
fundamental--adjoint action at finite temperature in more detail. We show
that the first order bulk transition in pure gauge SU(3) in the
fundamental--adjoint coupling plane indeed separates from the thermal
deconfinement transition provided $N_t$ is made large enough.

According to universality arguments \cite{Pisarski84}, QCD with only two
massless quarks has a second order thermal deconfining phase transition,
which becomes a smooth crossover if the quarks are massive.
Lattice QCD simulations with two light flavors of staggered quarks support
this scenario. However, with two flavors of Wilson quarks recent
simulations indicate that at large values of the hopping parameter the
phase diagram is more complicated \cite{wilson_woes}: for $N_t=6$ and at
hopping parameter $\kappa \sim 0.19$ the system has a strong first
order--like transition, where the plaquette expectation value has a sharp
discontinuous jump. However, the average Polyakov loop remains small at
this point, and starts to increase only when $\beta$ is considerably
larger. This behavior is strongly reminiscent of separated bulk and thermal
transitions, as in the pure gauge SU(3) model with mixed
fundamental--adjoint action. In fact, these systems resemble each other so
much that one is tempted to assume that the {\em dynamical\,} reason for
these transitions is the same: the dynamical Wilson fermions induce a
strong adjoint term in the effective pure gauge action. We therefore set
out to measure such an induced adjoint term employing a microcanonical
demon method.

\section{The fundamental--adjoint action}
\label{eqofm}

Let us consider the action $S$ for SU($N$) lattice gauge theory with
fundamental and adjoint coupling:
\be
S= \beta_f \sum_{P} [1 - \frac{1}{N} {\rm Re} {\rm Tr}_f U_{P} ] +
\beta_a \sum_{P} [1 - \frac{1}{N^2} {\rm Tr}_f U^\dagger_{P} {\rm Tr}_f
U_{P}].
\label{aaction}
\ee
Here ${\rm Tr}_f = {\rm Tr}$ is the trace in the fundamental
representation, $U_{P}$ is the path ordered product of the link matrices
along the elementary plaquette $P$. Using ${\rm Tr}_a U = |{\rm Tr}_f U|^2
- 1$ for the trace in the adjoint representation, the parametrization
(\ref{aaction}) is connected to the alternative one, used for general
one-plaquette actions \cite{GAKA},
\be
S= \beta_f \sum_{P} [1 - \frac{1}{N} {\rm Re} {\rm Tr}_f U_{P} ] +
\beta_a^\prime \sum_{P} [1 - \frac{1}{N^2-1} {\rm Tr}_a U_{P}]
\ee
by
\be
\beta_a = \frac{N^2}{N^2-1} \beta_a^\prime .
\ee

The $\Lambda$-parameters of the standard Wilson and the fundamental--adjoint
action are connected by~\cite{GAKA}
\be
\log \frac{\Lambda^W}{\Lambda^{FA}} = \frac{N^2+1}{8b_0} \left(
\frac{ \beta_a }{\beta_f+2\beta_a} \right)
\ee
with $b_0$ the one-loop coefficient of the $\beta$-function. Therefore, the
equivalent Wilson coupling, in perturbation theory, is
\be
\beta_W = \beta_f + 2\beta_a - \frac{N^2+1}{2}
\frac{\beta_a}{\beta_f+2\beta_a} ,
\label{Wil_coup}
\ee
and lines of constant $\beta_W$ in (\ref{Wil_coup}) should represent lines
of constant physics (up to two loop and non-perturbative corrections).
Of course, at the relatively large couplings that we will be working with,
the relation (\ref{Wil_coup}) is not very reliable. We shall try to improve
the relation by using tadpole improved perturbation theory \cite{L_M},
where a factor $u_0$, with $u_0^4 = {\rm Tr} U_P / N$, is taken out of
every link field $U_\mu$. In the relation (\ref{Wil_coup}) tadpole
improvement amounts to the replacement $\beta_a \rightarrow u_0^4
\beta_a$ with $u_0$ computed in the equivalent Wilson theory. Compensating
for the $u_0^4$ as computed in perturbation theory then gives, instead
\be
\beta_W = \beta_f + 2\beta_a u_0^4 -
\frac{\beta_a u_0^4}{\beta_f+2\beta_a u_0^4} .
\label{Wil_coup_tp}
\ee

\section{The phase diagram in the fundamental--adjoint coupling plane}
\label{phase_diag}

\begin{figure}[htb]
\begin{center}
\leavevmode
\epsfysize=360pt
%\epsfbox{plqf_ba2p25.eps}
\epsfbox[90 40 580 490]{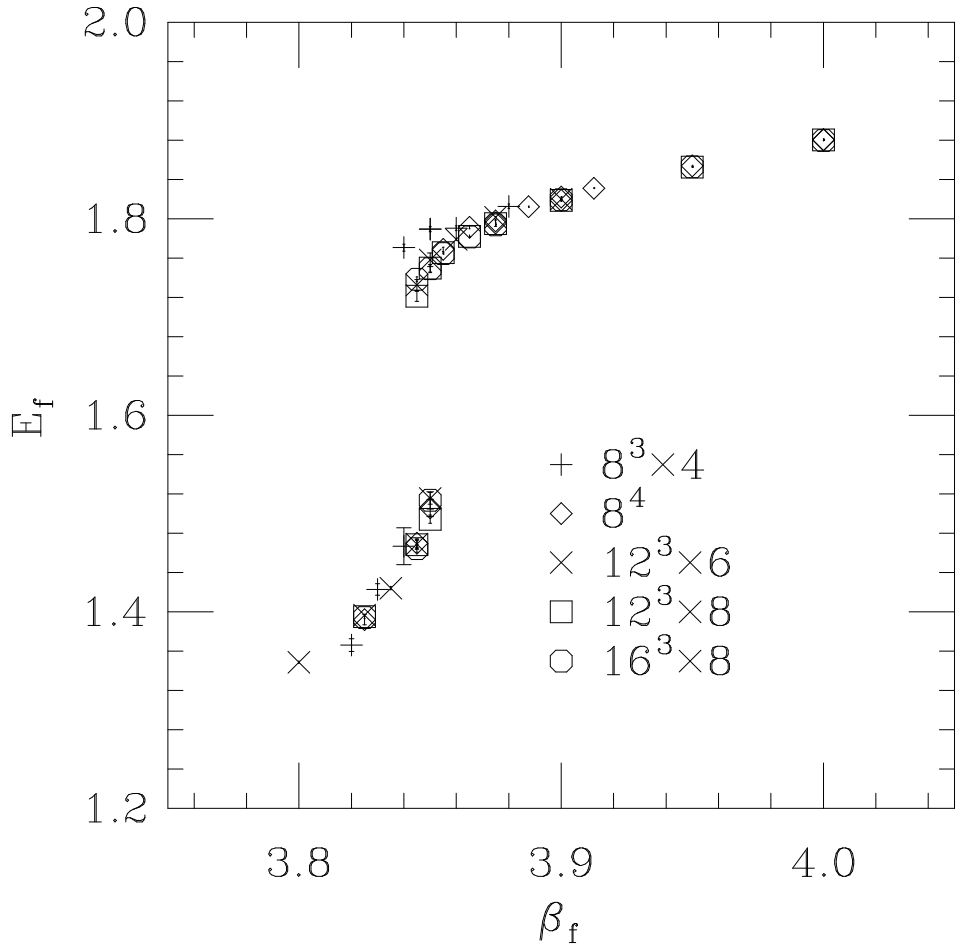}
%\epsffile{plqf_ba2p25.eps}
\end{center}
\caption{$E_f$ as function of $\beta_f$ for various lattice sizes, as
         indicated, at $\beta_a=2.25$.}
\label{plqf_ba2p25}
\end{figure}

Monte Carlo simulations with the mixed action described above were carried
out using a standard Metropolis updating scheme. First we found it
necessary to repeat the early calculation of Bhanot \cite{Bhanot82} to
locate the endpoint of the bulk phase transition more accurately. This was
done in a series of simulations with $8^4$ lattices by measuring the
discontinuity in the average fundamental plaquette $E_f = \langle{\rm Tr}
U_P\rangle$ and adjoint plaquette $E_a = \langle|{\rm Tr} U_P|^2/3\rangle$
along the first order line. Results are shown in Table~\ref{gaps}. To
support the identification as a bulk transition Fig.~\ref{plqf_ba2p25}
shows $E_f$, obtained from various lattice sizes, as function of $\beta_f$
for $\beta_a=2.25$ We note that $E_f$ jumps for all lattices at the same
point.

\begin{table}[htb]
\begin{center}
\begin{tabular}{|l|l|l|l|} \hline
$\beta_a$ & $\beta_f$ & $\Delta E_f$ & $\Delta E_a$ \\ \hline\hline
 3.0  & 3.27 & 0.656( 2) & 0.464(1) \\ \hline
 2.75 & 3.47 & 0.547( 2) & 0.398(2) \\ \hline
 2.5  & 3.66 & 0.424( 5) & 0.315(4) \\ \hline
 2.35 & 3.77 & 0.336( 9) & 0.250(7) \\ \hline
 2.25 & 3.85 & 0.252(10) & 0.193(8) \\ \hline
\end{tabular}
\end{center}
\caption{The jumps in in the average plaquette and average adjoint
         plaquette values across the bulk transition line.}
\label{gaps}
\medskip\noindent
\end{table}

The location of the phase transitions for the different values of $\beta_a$
considered was first obtained with runs from mixed starts. At the two
$\beta_f$ values that straddle the transition we then made runs  with hot
and cold starts to obtain the gaps in Table~\ref{gaps}. When we found
metastable states in both cases we took the larger gap measured. In any
case, the errors listed in Table~\ref{gaps} are statistical only. We
estimate systematic errors to be of order 0.010 to 0.020 in $\Delta E_f$.

\begin{figure}[htb]
\begin{center}
\leavevmode
\epsfysize=360pt
%\epsfbox{gap_fit.eps}
\epsfbox[90 40 580 490]{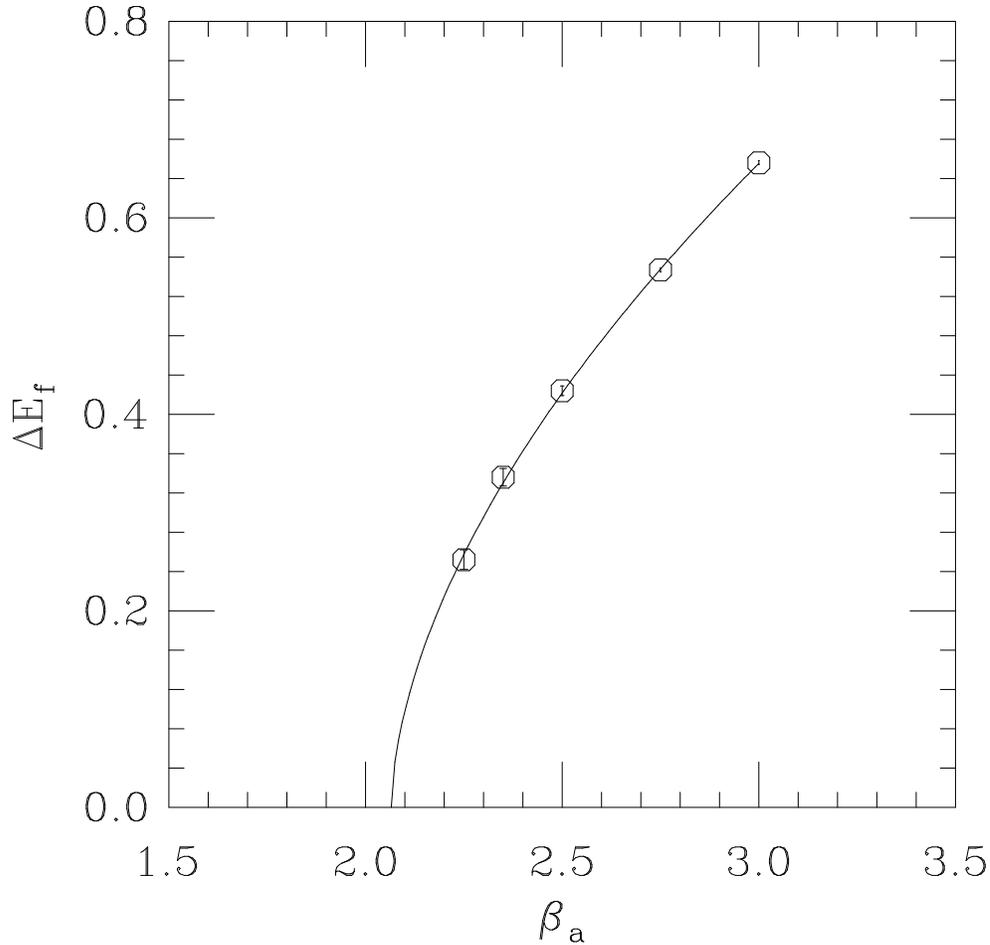}
%\epsffile{gap_fit.eps}
\end{center}
\caption{The average plaquette gap $\Delta E_f$ as function of $\beta_a$
         together with the fit to determine the endpoint $\beta_a^*$.}
\label{gap_fit}
\end{figure}

The endpoint of the bulk transition line $(\beta^*_f,\beta^*_a)$ is then
estimated by fitting to
\be
\Delta E_f = c ~ (\beta_a - \beta_a^*)^p .
\ee
Such a fit works very well, as shown in Fig.~\ref{gap_fit}, having $\chi^2
= 1.16$ for 2 degrees of freedom. In this region of the parameter space the
bulk transition line is essentially straight, and from the endpoint value
$\beta_a^* = 2.06(8)$ we infer the location of the endpoint of the bulk
transition line as
\be
     (\beta^*_f,\beta^*_a) = (4.00(7), 2.06(8)),
\ee
shown with our improved location of the bulk phase transition line in
Fig.~\ref{phas_diag_T0}. The error of 0.08 in $\beta_a^*$ includes our
estimate of the systematic uncertainties in the gaps, resulting as
mentioned above from not obtaining them exactly at the critical coupling.
The statistical error from the fit to the data in Table~\ref{gaps} is only
0.03. The best fit values for the other parameters are $c=0.68(1)$ and
$p=0.58(3)$.

\section{Results at nonzero temperature}

As in pure SU(2), the SU(3) pure gauge deconfinement transition extends to
non-zero $\beta_a$ values, since the global Z(3) symmetry is not affected
by the addition of an adjoint coupling. As we increase the adjoint coupling
the transitions for different $N_t$ move closer together, and closer to the
so called crossover region from strong to weak coupling behavior, which can
be regarded as an extension of the bulk phase transition line.

\begin{figure}[htb]
\begin{center}
\vskip 30mm
\leavevmode
\epsfysize=360pt
%\epsfbox{phas_diag_T.eps}
\epsfbox[90 40 580 490]{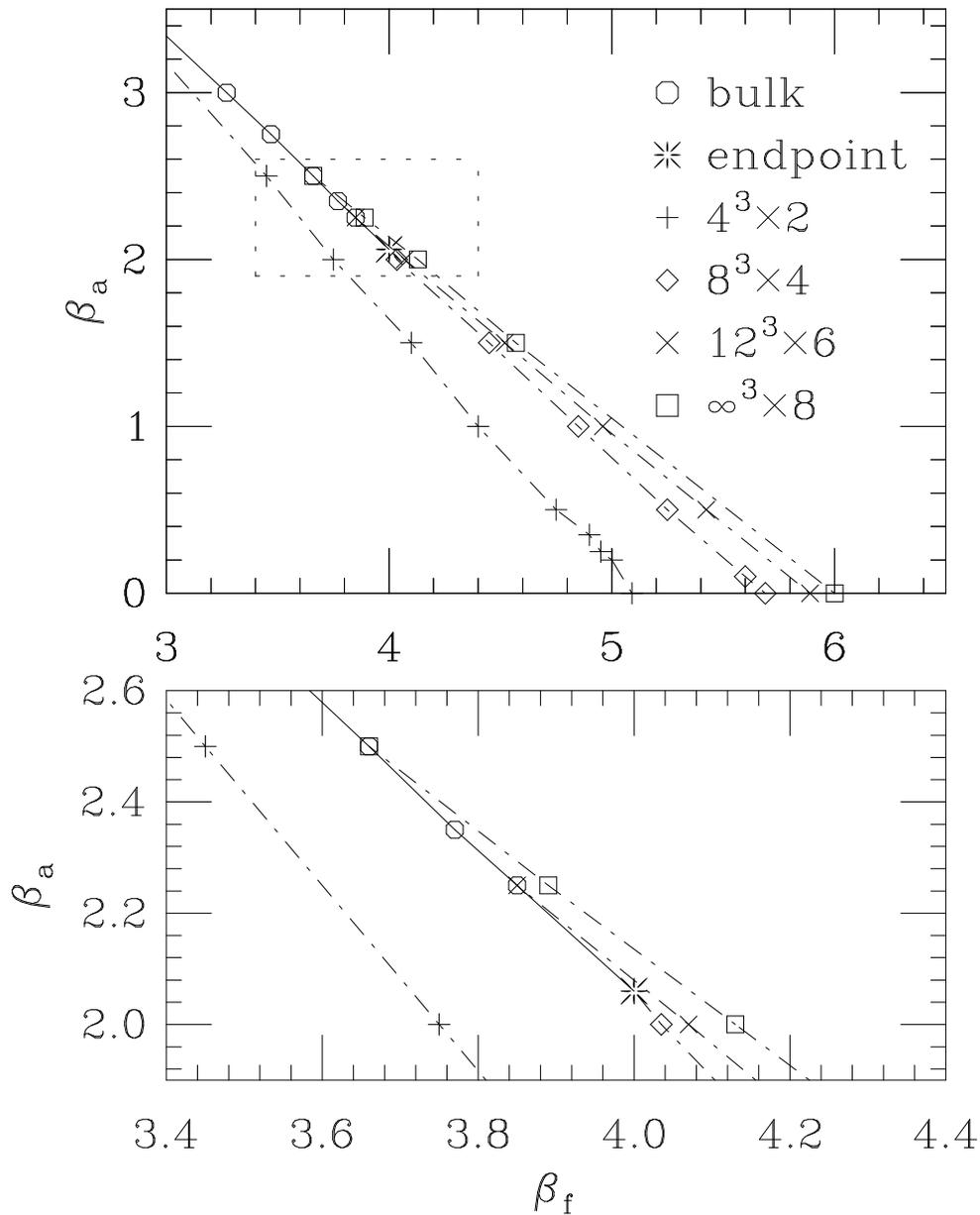}
%\epsffile{phas_diag_T.eps}
\end{center}
\caption{The phase diagram together with the thermal deconfinement
         transition points for $N_t=2$, 4, 6 and 8. The lower plot shows
         an enlargement of the region around the end point of the bulk
         transition.}
\label{phas_diag_T}
\end{figure}

To explore the high temperature behavior of the mixed action theory, we
studied the deconfinement transition on lattices with temporal extent
$N_t=2$, 4, 6 and 8. Note that in the first 3 cases we limited ourselves to
lattices with aspect ratio $N_s/N_t = 2$. That proved sufficient to find
the deconfinement transitions with sufficient accuracy. The thermal
deconfinement transition point is defined as the location of maximum slope
in the modulus of the Polyakov loop $\langle |P| \rangle$. All transition
points we obtained are shown in Fig.~\ref{phas_diag_T} and
Table~\ref{FT_phase_transition}.  For $N_t=2$ the $\beta_a = 0$
deconfinement transition continues smoothly into the bulk transition, which
is shifted significantly from its large $N_t$ value, as $\beta_a$
increases. For $N_t = 4$ at $\beta_a = 2.0$ the deconfinement phase
transition joins and is indistinguishable from from the bulk transition,
just as was found for SU(2) \cite{Gavai94}. But for $N_t = 6$ at $\beta_a =
2.0$, there is a clear separation, with the bulk crossover occurring at
$\beta_f = 4.035(5)$ and the thermal deconfinement transition at $\beta_f =
4.07(2)$. At $\beta_a=2.25$, even for the $12^3 \times 6$ lattice, bulk and
deconfinement transition coincide, both occurring at $\beta_f=3.850(5)$.

\begin{table}[htb]
\begin{center}
\begin{tabular}{|l|r|l|r|l|} \hline
   \multicolumn{1}{|c|}{ $\beta_a$}&
   \multicolumn{4}{c|}{$\beta_f$}\\
 \cline{2-5}
  & $N_t=2$ & $N_t=4$ &$N_t=6$ &$N_t=8$  \\ \hline\hline
 0.0  & 5.0941(4) & 5.6925(2) & 5.8941(5) & 6.001(25) \\ \hline
 0.5  & 4.75(5)   & 5.25(5)   & 5.425(20) & -         \\ \hline
 1.0  & 4.4(1)    & 4.85(5)   & 4.96(2)   & -         \\ \hline
 1.5  & 4.1(1)    & 4.45(5)   & 4.525(20) & 4.58(2)   \\ \hline
 2.0  & 3.75(5)   & 4.035(5)  & 4.07(2)   & 4.135(10) \\ \hline
 2.25 & -         & 3.845(5)  & 3.850(5)  & 3.89(1)   \\ \hline
 2.5  & 3.45(5)   & -         & -         & 3.660(5)  \\ \hline
\end{tabular}
\end{center}
\caption{The deconfinement transition points.}
\label{FT_phase_transition}
\medskip\noindent
\end{table}

\begin{figure}[htb]
\begin{center}
\leavevmode
\epsfysize=360pt
%\epsfbox{ModP_extr.eps}
\epsfbox[90 40 580 490]{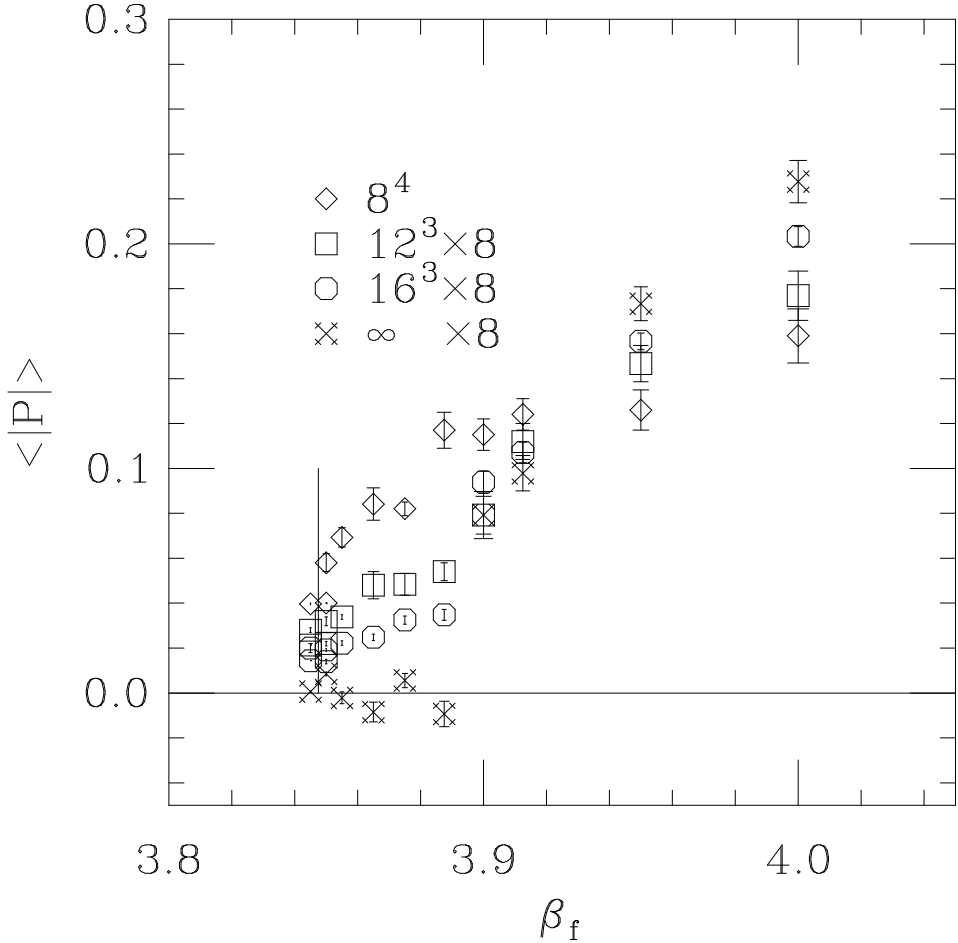}
%\epsffile{ModP_extr.eps}
\end{center}
\caption{$\langle |P| \rangle$ at $\beta_a = 2.25$ for $N_t=8$ and $N_s=8$,
 12 and 16 and the extrapolation to $N_s = \infty$.}
\label{ModP_extr}
\end{figure}

At larger $N_t$ the small size of the order parameter required additional
care in locating the phase transition. We carried out a finite size
analysis with simulations on $N_s^3\times 8$ lattices with $N_s = 8$,12,16.
In the confined phase $\langle |P| \rangle$ should vanish as $1/\sqrt{V_s}$
with increasing spatial volume, while in the deconfined phase it should
extrapolate to a nonzero value. To treat both cases identical we made fits
to $\langle |P| \rangle = P_\infty + c/\sqrt{V_s}$. The finite size
analysis is shown for $\beta_a=2.25$ in Fig.~\ref{ModP_extr}. Plots for
other $\beta_a$ are very similar. In this way we place the $N_t=8$
deconfinement transition for $\beta_a=2.0$ at $\beta_f=4.135(15)$, clearly
separated from the bulk crossover as well as from the $N_t=6$ deconfinement
transition. For $\beta_a=2.25$ the deconfinement transition occurs at
$\beta_f=3.89(1)$, still separated from the bulk transition at
$\beta_f=3.850(5)$, whereas at $\beta_a = 2.50$, no clear separation is
visible.

We conclude that with increasing $N_t$ the deconfinement transition line is
displaced toward weaker coupling, joining onto the bulk transition at larger
and larger values of $\beta_a$, consistent with universality. However,
trying to see low temperature continuum physics at larger values of the
adjoint coupling $\beta_a$ requires larger lattices to avoid strong
violations of asymptotic scaling associated with the bulk transition line.

\begin{figure}[htb]
\begin{center}
\leavevmode
\epsfysize=360pt
%\epsfbox{const_phys.eps}
\epsfbox[90 40 580 490]{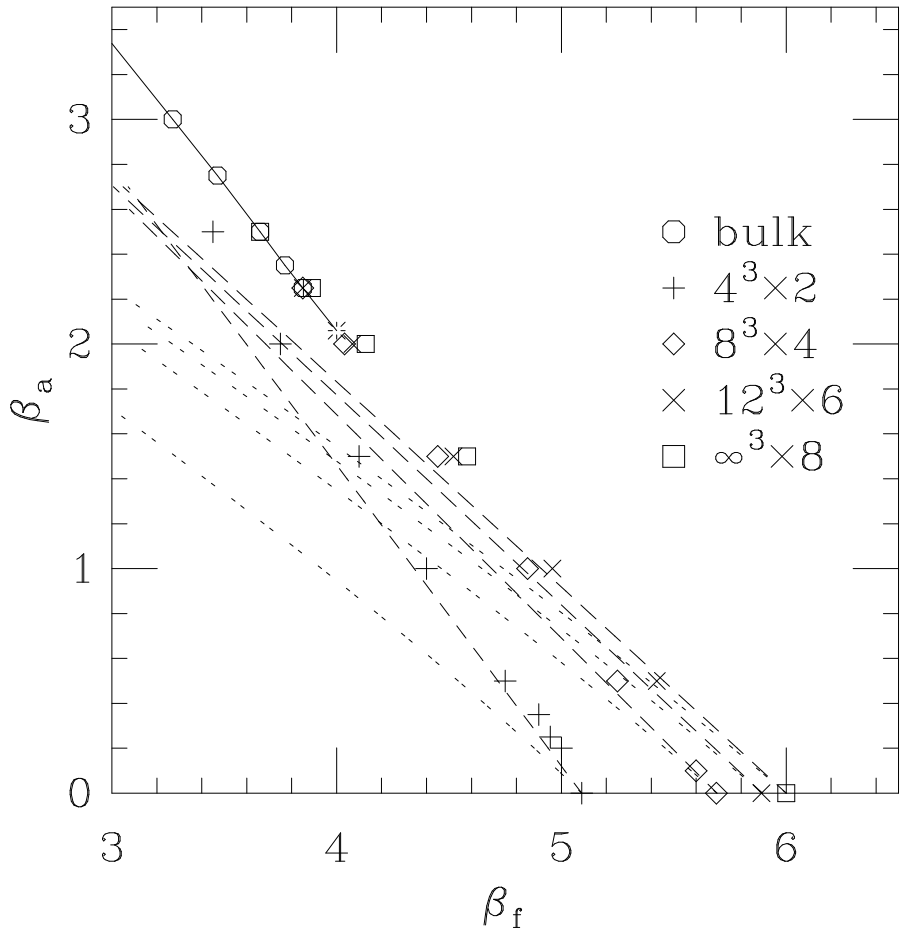}
%\epsffile{const_phys.eps}
\end{center}
\caption{Lines of constant physics as predicted by perturbation theory
         (dotted lines) and tadpole improved perturbation theory (dashed
         lines) together with the deconfinement transitions for $N_t=2$,
         4, 6, and 8.}
\label{const_phys}
\end{figure}

Having found non-perturbatively lines of constant physics, here defined as
lines of the thermal deconfinement transition for fixed $N_t$, we can
compare them to the perturbative predictions obtained from
eq.~(\ref{Wil_coup}). Not surprisingly, the perturbative prediction is
poor. Using, instead, the tadpole improved prediction
eq.~(\ref{Wil_coup_tp}) changes the perturbative result in the right
direction, but still fails to give the correct lines of constant physics,
as shown in Fig.~\ref{const_phys}. It is interesting to notice, though,
that $E_f = \langle {\rm Tr} U_P \rangle$ is almost constant along the
deconfinement transition lines. Therefore, equal effective coupling
$\beta_{\rm eff} = 2/(1-E_f/3)$ describes lines of constant physics remarkably
well.

\section{Couplings induced by the fermionic determinant}

Full QCD simulations with two flavors of staggered fermions support the
view that there is no thermal phase transition at positive quark masses,
and a second order transition at zero quark mass \cite{Karsch93} (a
scenario also supported by universality arguments \cite{Pisarski84}).
However, the phase diagram with dynamical Wilson fermion has turned out to
be complicated for large values of the hopping parameter. It was found that
for $N_t = 6$ and hopping parameter $\kappa \sim 0.19$, there is an
apparent first order transition leading to a jump in the average plaquette.
At slightly higher $\beta_f$ the Polyakov loop expectation value $\langle
{\rm Re} P \rangle$ remains small in a manner remarkably similar to our
findings for the region separating the bulk and thermal phase transition in
the mixed action pure gauge theory. It is tempting to speculate that the
first order phase transition seen with Wilson quarks is a bulk transition
related to the first order pure gauge adjoint bulk transition.

It has been shown, that the location of the deconfinement transition with
staggered fermions can be surprisingly well explained by a change in the
fundamental coupling induced by heavy fermions \cite{Hasenfratz94}. Let us
see how new terms in the gauge action are induced by Wilson fermions. The
discretized action $S_0$ for $n_f$ flavors of Wilson fermions is
\be
S_0 = S_g + \sum_{f=1}^{n_f} \sum_{n,m} {\bar\psi}_n^f M_{nm}[U] \psi_m^f,
\label{fermionact}
\ee
where
\be
M_{nm}[U]=\delta_{nm}-\kappa\sum_\mu((1-\gamma_\mu)U_{n\mu}\delta_{n+\mu,m}
+(1+\gamma_\mu)U^{\dagger}_{n\mu}\delta_{n-\mu,m})
\ee
and $S_g$ is the pure gauge action. The fermions can be integrated out and
the effective gauge action $S_{\rm eff}$ can be obtained by a hopping
parameter ($\kappa$) expansion.
\ba
S_{\rm eff}&&=S_g - n_f {\rm Tr} \ln M[U] \cr
&&=S_g+ n_f \sum_{C}\kappa^{l[C]}{\frac{1}{l[C]}}
{\rm Tr}\left (\prod_C(1\pm\gamma_\mu)\right )\cdot
\left ({\rm Tr} U_C+{\rm Tr} U^{\dagger}_C\right ),
\ea
where the sum is over all (unoriented) closed loops $C$ and $l[C]$ is the
length of the loop. For small hopping parameters $\kappa$, the shift for
the fundamental coupling comes from loops around a single plaquette and is
given by $\Delta\beta_f = n_f {\frac{4}{3}} \kappa^4$. The shift in the
adjoint coupling comes from closed loops winding three times around a
plaquette. For small $\kappa$ it is $\Delta\beta_a \propto \kappa^{12}$.
Similarly the staggered fermion action induces an adjoint coupling. Indeed
with eight flavors, a bulk transition has been found~\cite{columbia_nf8},
but studies have not yet been done to determine whether there is evidence
for a similar separation of the bulk transition and the thermal crossover.

The hopping parameter expansion outlined above is accurate only when
$\kappa$ is very small. Results with much broader validity range can be
obtained with the heavy quark perturbation method. This method was used by
Hasenfratz and DeGrand \cite{Hasenfratz94} to calculate $\Delta\beta_f$.
They found shifts $\Delta\beta_f$ that agreed very well with MC data.
However, such a heavy quark perturbation calculation has not been performed
for $\Delta\beta_a$.

\subsection{Demon Algorithm}

We use the microcanonical demon method \cite{Creutz84,Yee94,Hasenbusch94}
to project out the induced effective gauge action from full fermionic
simulations.  Our ansatz for the effective action is given by
\eq~(\ref{aaction}), with a priori unknown coupling constants $\beta_i =
\beta_f$ or $\beta_a$. For both of the coupling constants we introduce a
{\em demon\,}, which is a real-valued action variable with $0 \le D_i
\le D_i^{\rm max}$.
Effectively, the fermionic action is used as a heat bath to thermalize
the demons: starting from a configuration generated with the original
action, we update the system microcanonically with demons.  During the
microcanonical update $S_i + D_i = {\rm const.}$ separately for both
fundamental and adjoint parts of the action.  After the microcanonical
update, the old gauge configuration is discarded, and the demon update
is started again from a new gauge configuration, while preserving the
demon values.  The principle of the method is shown graphically in
\fig\ref{figalgo}.

\newcommand{\conf}{\circle{2}}
\newcommand{\demon}{\circle*{2}}
\newcommand{\txt}{\small\it}
\begin{figure}[h]
\setlength{\unitlength}{1mm}
\centerline{
\begin{picture}(75,45)
%\thicklines
\put(3,40){\vector(1,0){9}}
\put(3,5){\vector(1,0){9}}
\put(15,40){\conf}
\put(15,37){\vector(0,-1){11}}\put(17,32){\makebox(0,0)[l]{\txt copy}}
\put(15,23){\conf}
\put(15,13){\vector(0,-1){5}}\put(18,16){\makebox(0,0)[l]{\txt microcanonical}}
\put(15,13){\vector(0,1){8}}\put(18,12){\makebox(0,0)[l]{\txt update}}
\put(15,5){\demon}
\put(13,3){\dashbox(4,22)[br]}
\put(18,40){\vector(1,0){29}}
		\put(33,43){\makebox(0,0){\txt update with $S_0$}}
\put(18,5) {\vector(1,0){29}} \put(33,2) {\makebox(0,0){\txt copy}}
\put(18,23){\vector(1,0){5}}  \put(25,23){\makebox(0,0)[l]{\txt discard}}
%        \put(21,23){\line(1,0){2}}
%	\put(23,20.5){\line(0,1){5}}
%	\put(23.7,21.25){\line(0,1){3.5}}
%	\put(24.4,22){\line(0,1){2}}
\put(50,40){\conf}
\put(50,37){\vector(0,-1){11}}
\put(50,23){\conf}
\put(50,13){\vector(0,-1){5}}
\put(50,13){\vector(0,1){8}}
\put(50,5){\demon}
\put(53,40){\vector(1,0){9}}
\put(53,5){\vector(1,0){9}}
\put(60,18){\conf} \put(63,18){\makebox(0,0)[l]{\txt gauge configurations}}
\put(60,14){\demon}\put(63,14){\makebox(0,0)[l]{\txt demons}}
\end{picture}}
\caption[0]{Flow diagram of the demon algorithm.  The configurations
            are generated with the original fermionic action $S_0$
	    (top) \eq\nr{fermionact}, and copied for the
	    microcanonical demon update.  At the end of the
	    microcanonical update, the demon values are copied for the
	    next update phase (bottom).}
\label{figalgo}
\end{figure}
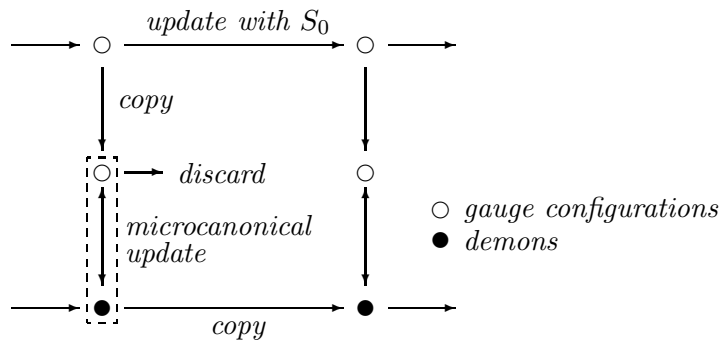

The microcanonical update phase is performed with a Metropolis
algorithm: the proposed new gauge link is accepted only if the demons
can give or take the amount of energy needed:
\be
 \Delta S_i \ge D_i \h \mbox{and} \h -\Delta S_i \le D_i^{\rm max} - D_i
\ee
for both $i=f$ or $a$.  Because both of the demons have to accept the
update step for it to be performed, the demon distributions are
correlated.  After the whole update procedure is repeated many times, the
demons attain an equilibrium distribution and we can measure $\langle
D_i\rangle$.  The induced effective couplings can be solved from
equations
\be
\langle D_i \rangle=\frac{1}{\beta_i} -
\frac{D_i^{\rm max}}{\exp(\beta_i D_i^{\rm max}) - 1}\, .
\la{dembeta}
\ee
It is not necessary to limit the demon action from above if the
expected value of the corresponding coupling constant is large.  In
our case, we had to limit only the adjoint demon values from above;
for the fundamental demon $D_f^{\rm max} = \infty$ and in this case
\eq\nr{dembeta} reduces to $\beta_f = 1/\langle D_f\rangle$.

One drawback of the demon method is that the final results can depend
on the details of the update procedure.  Some of the factors affecting
the results are the range and shape of the distribution from which the
new gauge matrix is chosen in the Metropolis update (true even if the
detailed balance is always satisfied), the acceptance rate, and the
`length' of the demon update phase: one can update each configuration
microcanonically until the demons and the systems are properly
thermalized, or one can stop the update after only one -- or even
partial -- update sweep.  These effects are easy to observe in some
simple test models where the energy distributions can be exactly
calculated.  This does not mean that some methods are correct, some
incorrect; different methods only correspond to different projections
from the original action to the functional space spanned by the ansatz
for the effective action.  Nevertheless, one should not ignore this
problem when using the demon method.  Furthermore, let us note that if
the effective action is completely equivalent to the original action,
the above effects vanish and the demon method yields a unique $\beta_i$.
Conversely, if the effective action is close to the true action --
which should be the case for all reasonable effective actions -- one
can expect that the differences between the methods will be small.

The autocorrelations between configurations can introduce further
systematic errors \cite{Hasenbusch94}.  These errors are unrelated to
the errors above.  However, in most cases the differences are expected
to behave as $1/V$ as $V\rightarrow\infty$ (the heat capacity of the
system $\gg$ the heat capacity of the demons).

In our case we have a dramatically truncated effective action, and a
priori we do not have any guarantee of the `goodness' of the ansatz.
The results were checked by using different update schemes: we
performed 1 or 20 trajectories with the fermionic action between the
microcanonical update phases, and 1/8, 1 or 50 microcanonical update
sweeps for each gauge configuration.  In all our tests the possible
differences were completely overwhelmed by the statistical noise,
indicating that the method is quite robust to the accuracy we reached.
The results given below are all calculated by performing one
microcanonical update sweep for each gauge configuration and after each
fermionic trajectory.

\subsection{Simulations and Results}

\begin{figure}[tb]
\begin{center}
\leavevmode
\epsfysize=360pt
%\epsfbox{run_loc.eps}
\epsfbox[90 260 540 650]{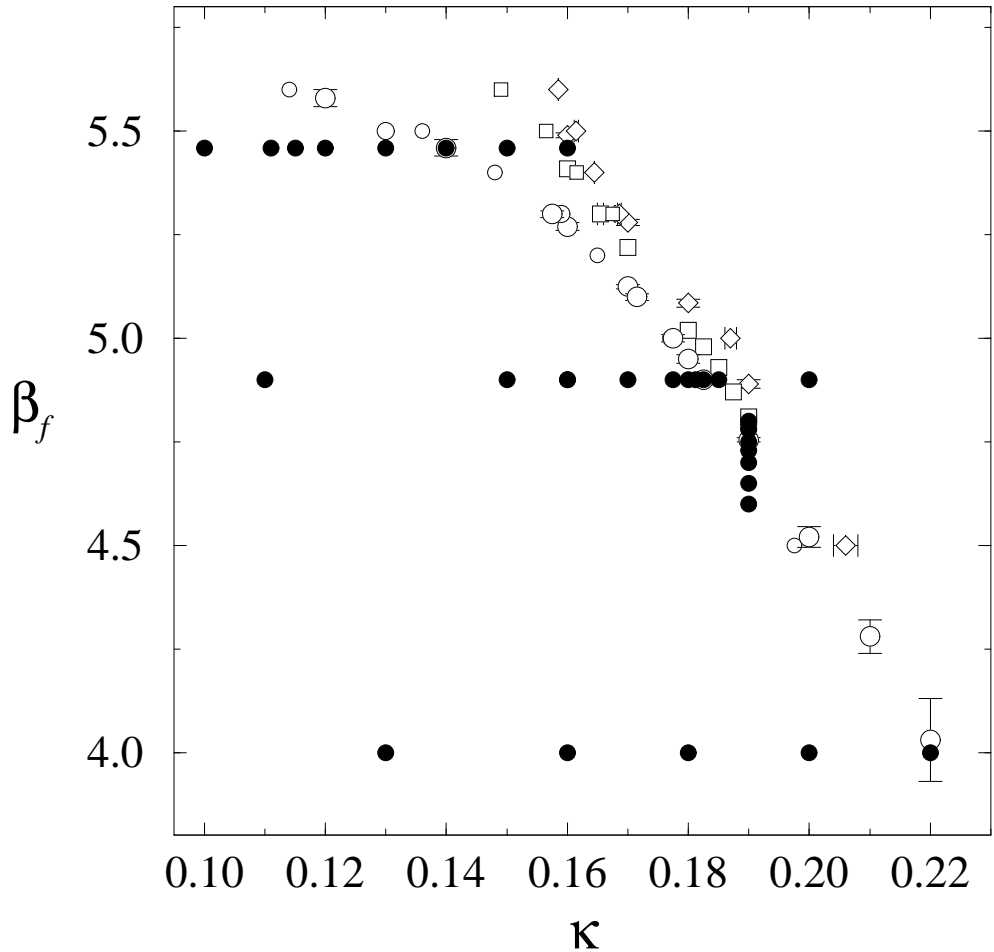}
%\epsffile{run_loc.eps}
\end{center}
\caption[0]{The location of the dynamical Wilson fermion runs in the
            $(\beta^0_f,\kappa)$-plane (filled circles).  The open
   	    circles indicate the location of the $N_t=4$ finite
   	    temperature crossover, the squares the $N_t=6$ phase
   	    transition, and the diamonds the zero-temperature
   	    $\kappa_c$.  The data is from ref.~\cite{wilson_woes} and
   	    from references therein.}
\label{run_loc}
\end{figure}

\begin{figure}[tb]
\begin{center}
\leavevmode
\epsfysize=360pt
%\epsfbox{bf_ind.eps}
\epsfbox[90 260 540 650]{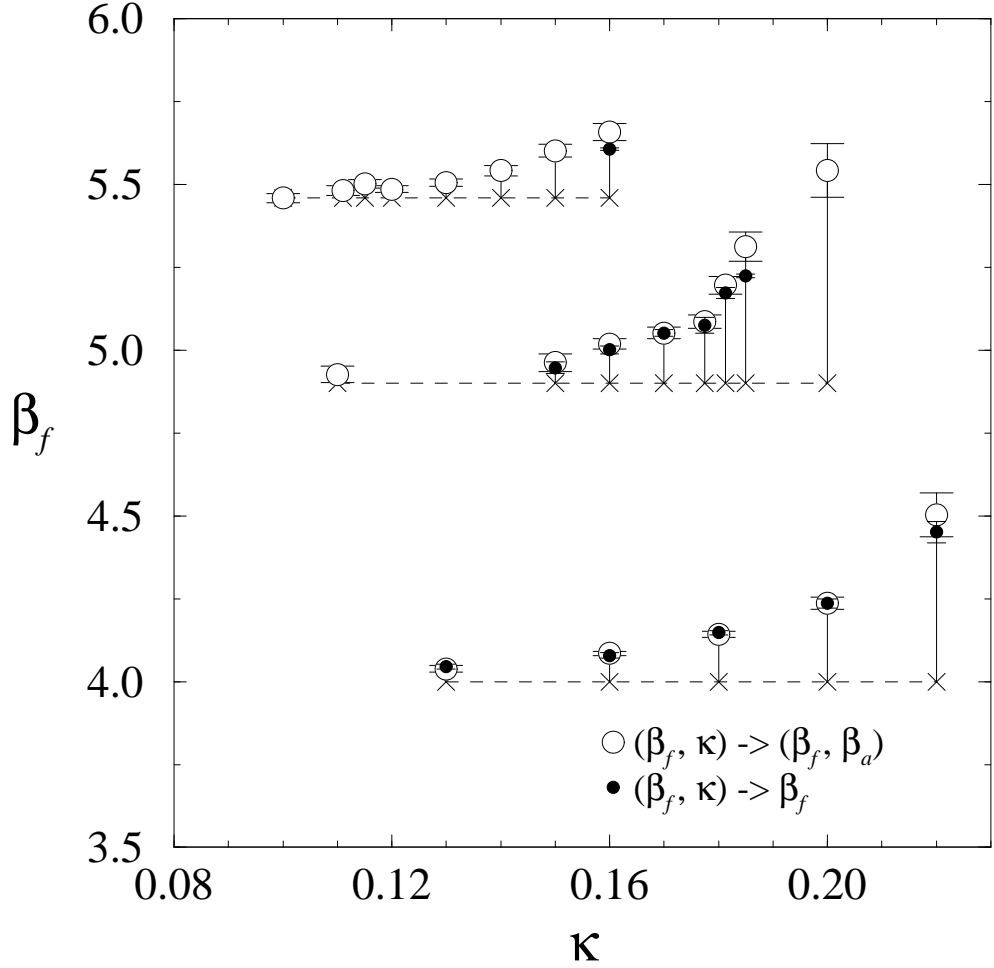}
%\epsffile{bf_ind.eps}
\end{center}
\caption{The induced $\beta_f$. Small crosses on horizontal dashed lines
         indicate the simulation $(\beta_f^0,\kappa)$, open circles
         and black dots the measured $\beta_f$-values using the
         effective fundamental--adjoint and fundamental only gauge
         action, respectively. The length of the vertical bars gives
         the magnitude of the induced $\Delta\beta_f$.}
\label{bf_ind}
\end{figure}

\begin{figure}[tb]
\begin{center}
\leavevmode
\epsfysize=360pt
%\epsfbox{ba_ind.eps}
\epsfbox[110 260 560 650]{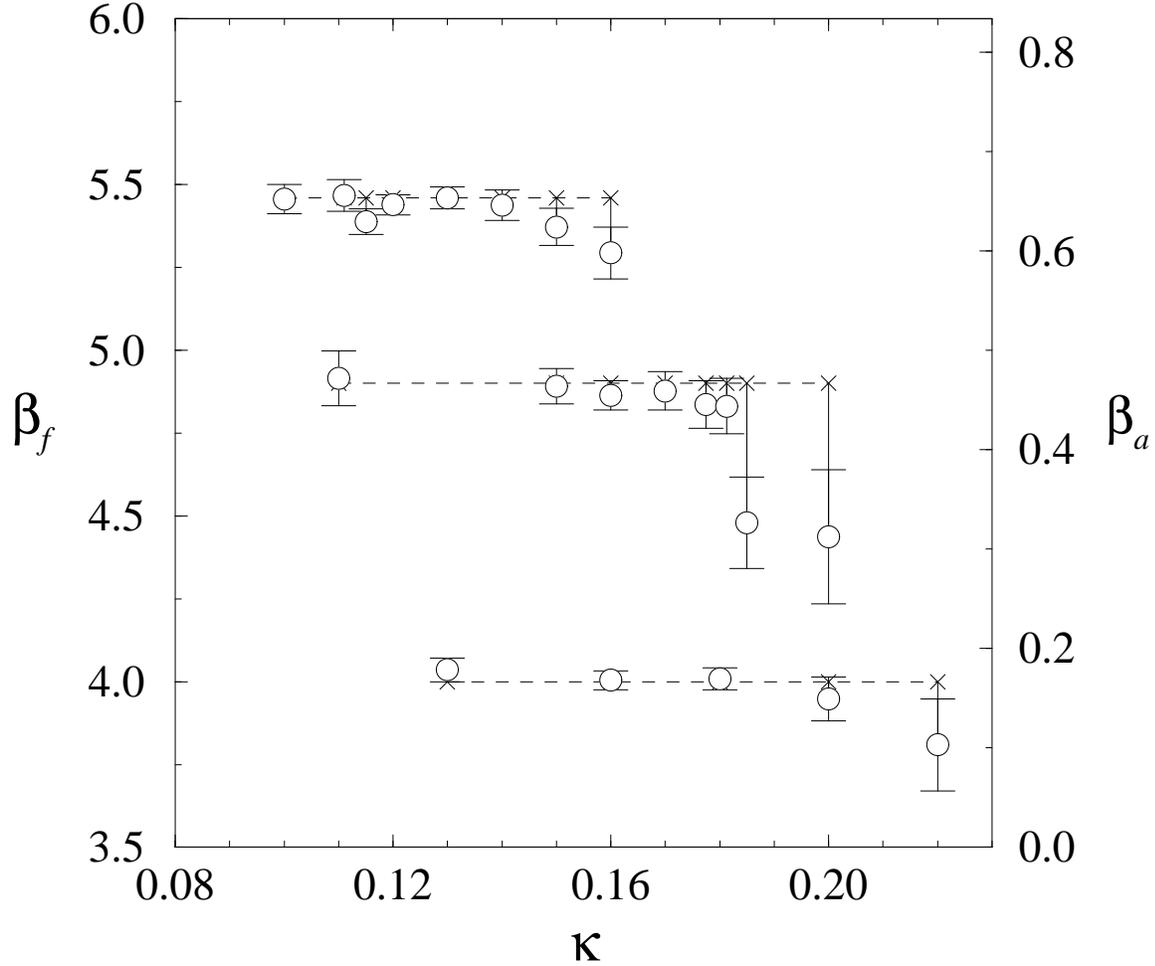}
%\epsffile{ba_ind.eps}
\end{center}
\caption{The induced $\beta_a$. The scale on the right gives the magnitude
         of the induced $\beta_a$, measured as the vertical distance
         between the plot symbols and dashed horizontal lines. Points
         below the dashed lines indicate negative $\beta_a$.}
\label{ba_ind}
\end{figure}

\begin{figure}[tb]
\begin{center}
\leavevmode
\epsfysize=360pt
%\epsfbox{k19_ind.eps}
\epsfbox[90 260 540 650]{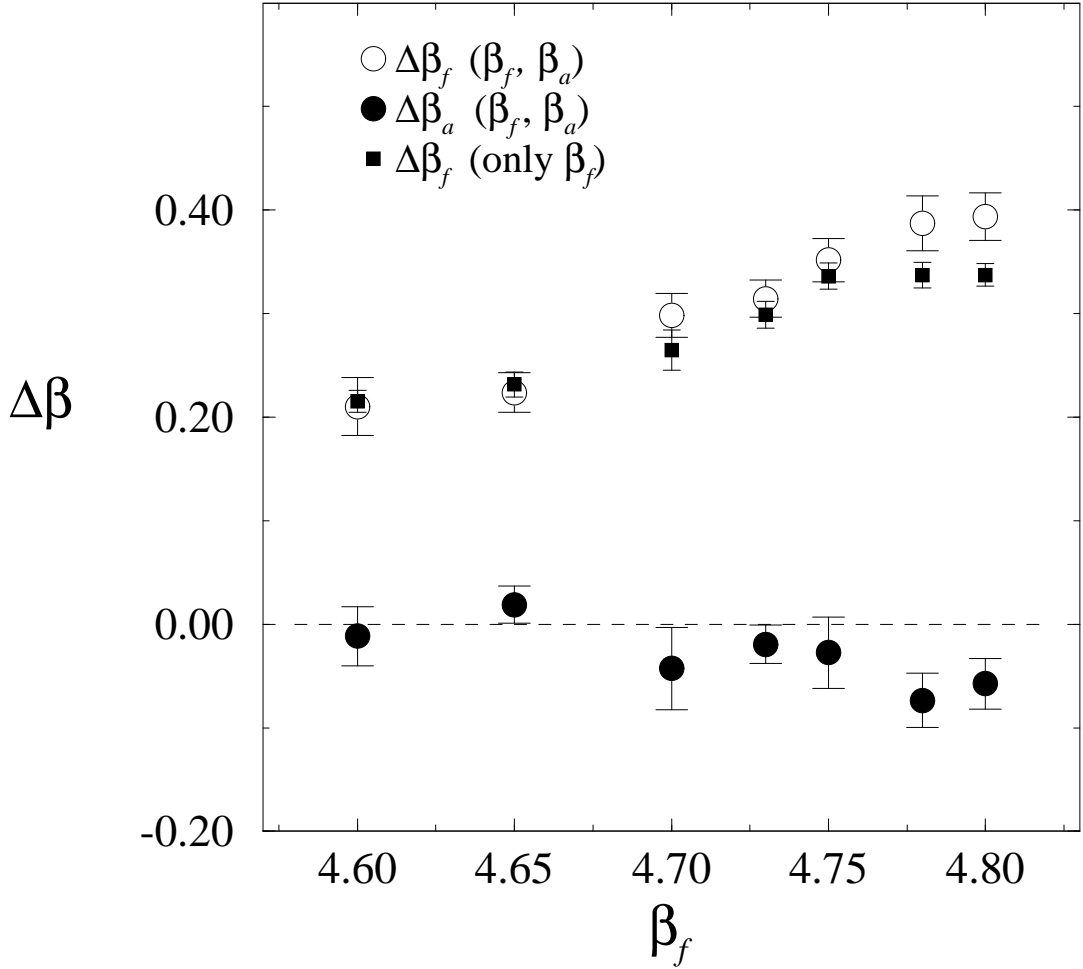}
%\epsffile{k19_ind.eps}
\end{center}
\caption{The induced $\beta_f$ and $\beta_a$, when $\kappa=0.19$.}
\label{k19_ind}
\end{figure}

\begin{figure}[tb]
\begin{center}
\leavevmode
\epsfysize=360pt
%\epsfbox{mq_bf.eps}
\epsfbox[90 260 540 650]{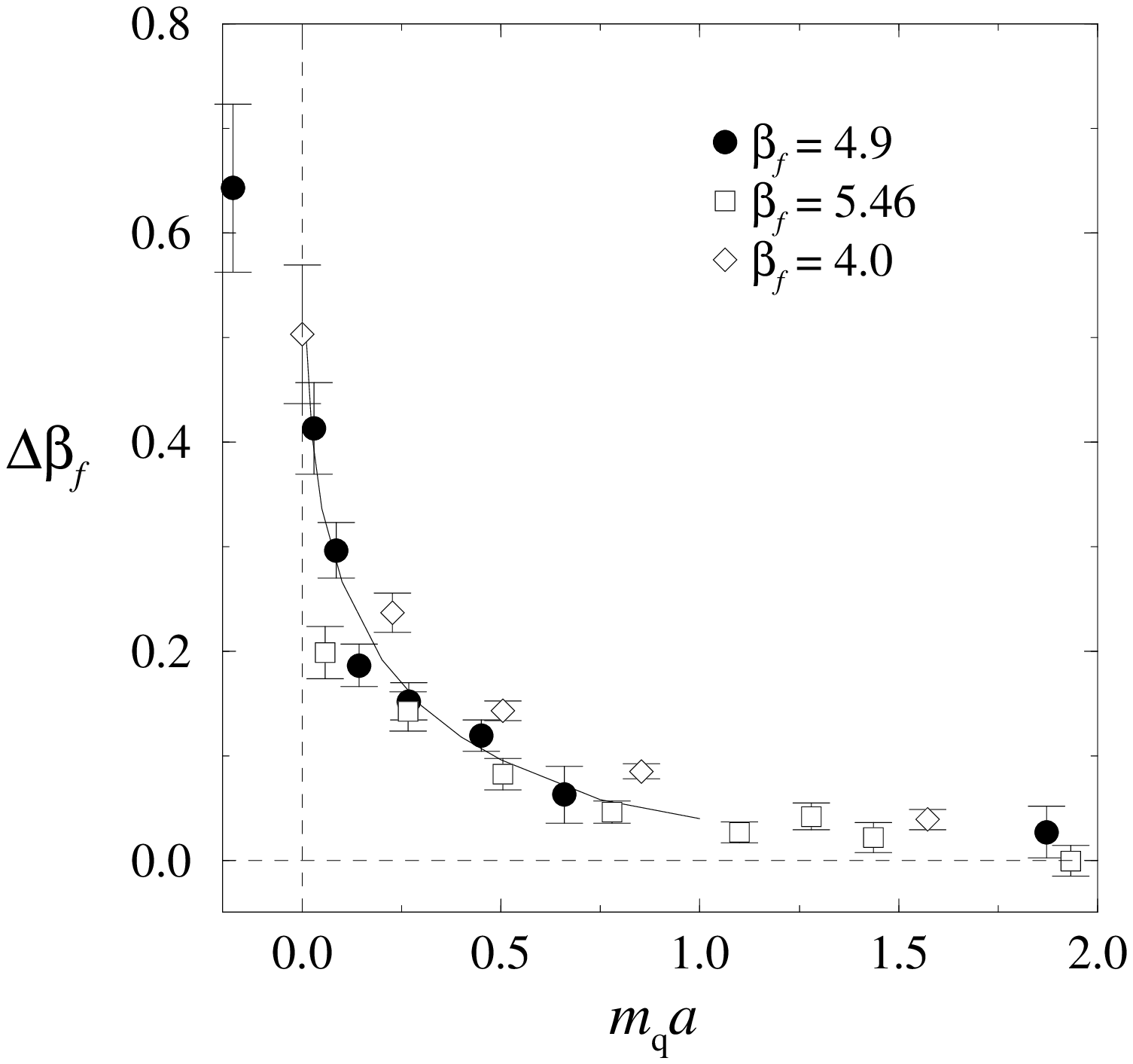}
%\epsffile{mq_bf.eps}
\end{center}
\caption{Comparison of the analytical result for $\Delta\beta_f$ \protect
         \cite{Hasenfratz94} by Hasenfratz and DeGrand (solid line) to the
         MC data.}
\label{mq_bf}
\end{figure}

We tested the demon method by applying it to pure gauge
fundamental--adjoint simulations. On a $4^4$-lattice we simulated the
system at couplings $(\beta_f,\beta_a)_0 = (3.6,1.8)$ and $(4.0,2.0)$. The
induced couplings were $(3.594(6),1.812(17))$ and $(4.017(27),1.92(6))$,
respectively, compatible with the input values. The latter coupling pair is
very close to the bulk transition line in the $(\beta_f,\beta_a)$-plane.
When we used an effective action consisting only of the $S_f$-part,
$\beta_f$ was 4.688(9) and 6.044(12); the latter value is very close to the
extension of the bulk transition line to the $\beta_a=0$ -axis.

If the Wilson fermion action induces a strong adjoint coupling giving rise
to a bulk fundamental--adjoint transition, one should be able to observe
the induced coupling already in small volumes. We performed simulations on
$4^4$ lattices with 28 different $(\beta^0_f,\kappa)$ pairs.
Fig.~\ref{run_loc} shows the location of all runs performed. In
Figs.~\ref{bf_ind} and \ref{ba_ind} we show the measured $\beta_f$ and
$\beta_a$ calculated with $\beta^0_f=4.0$, 4.9, 5.46, and several $\kappa$
values, and in Fig.~\ref{k19_ind} the results with constant $\kappa=0.19$.
In Figs.~\ref{bf_ind} and \ref{k19_ind} we also show $\beta_f$ when $S_{\rm
eff}=\beta_f S_f$ only.

When $\kappa$ is small, the quarks are very massive and the induced
couplings are quite small (left side of Figs.~\ref{bf_ind} and
\ref{ba_ind}). When $\kappa$ is increased, we approach the critical line
where $m_{\rm q}\rightarrow 0$, and the fermionic contribution to the
action becomes more significant. This is clearly visible as an increase in
$\beta_f$ in Figs.~\ref{bf_ind} and \ref{k19_ind}. The critical values of
$\kappa$ are approximately 0.16 ($\beta^0_f=5.46$), 0.19 (4.9) and 0.22
(4.0). We observe no significant increase in $\beta_a$; on the contrary,
when $\kappa_c$ is approached, the induced $\beta_a$ becomes slightly
negative!  The minor role of the adjoint action is also evident from the
fact that $\beta_f$ remains practically the same whether we use the
$S_a$-term of the effective action or not. We also checked the results with
a few simulations on $6^4$ lattices with similar results.

In Fig.~\ref{mq_bf} we compare $\Delta\beta_f$ to the predictions of
ref.~\cite{Hasenfratz94}, as a function of quark mass $m_{\rm q}a =
\kappa^{-1} - \kappa_c^{-1}$. The agreement is very good, especially when
$\beta_f^0 = 4.9$.

\section{Conclusions}
\label{conclusions}

We have shown that the fundamental--adjoint pure SU(3) gauge theory behaves
as expected near the first order bulk transition. The thermal deconfinement
transition lines join onto the bulk transition lines for larger and larger
$\beta_a$. For $N_t=2$ the thermal transition continues smoothly into the
bulk transition which is, however, shifted from its location for larger
$N_t$. The $N_t=4$ thermal transition line joins the bulk transition line
very close to the end point at $\beta_a \sim 2.0$, for $N_t=6$ at $\beta_a
\sim 2.25$ and for $N_t=8$ at $\beta_a \sim 2.5$. Before joining the bulk
transition line, the thermal transition line for a larger $N_t$ is to the
right (at larger $\beta_f$) than for a smaller $N_t$. This finding is in
agreement with what is expected from the usual universality picture of
lattice gauge theories.

We speculated that the bulk transition observed with Wilson fermions for
$N_t=6$ and large $\kappa$~\cite{wilson_woes}, that looked very reminiscent
of the situation seen in the pure gauge fundamental--adjoint action model,
might be caused by an induced adjoint coupling. However, in measurements
with the microcanonical demon method we did not observe that Wilson
fermions induce any significant adjoint term in the pure gauge effective
action, while the induced fundamental term is very well described by
analytical calculations. It is thus very improbable that the transition
observed in $N_t=6$ Wilson thermodynamics simulations could be explained by
the fundamental--adjoint pure gauge transition, and the real cause of this
phenomenon remains to be uncovered.

\subsubsection*{Acknowledgments}

This work was partly supported by the U.~S.~Dept. of Energy under grants
\#~DE-FG02-8SER40213, \#~DE-FG05-85ER250000, \#~DE-FG05-92ER40742,
and \#~DE-FG02-91ER40661, and the U.S.~National Science Foundation
under grants PHY9309458. LK wants to thank the Antti Wihuri foundation
for additional support. TB, CD, LK, KR and DT thank the Institute
for Theoretical Physics, Santa Barbara, where the dynamical Wilson
fermion part of this project was initiated. KR would like to thank
M. Hasenbusch for many useful discussions. Computations were carried
out on the UNIX clusters at the Supercomputer Computations Research
Institute at The Florida State University
and at the Pittsburgh Supercomputer Center,
using several workstations in parallel with PVM, on
IBM RS6000 workstations at the University of Utah and Indiana
University, and on SUN workstations at the University of Arizona and
Indiana University. Many of the simulations on the larger lattices were done
on the IBM SP1 at the Cornell Theory Center, which receives major
funding from the U.S.~National Science Foundation and the State of New
York, with additional funding from the Advanced Research Projects
Agency, the National Institute of Health, IBM Corporation, and other
members of the Cornell Theory Center's Corporate Research Institute.

\end{document}